# Measuring Communication Quality of Interest Rate Announcements[*]

Jonathan Benchimol,[1] Itamar Caspi[1] and Sophia Kazinnik[2]

June 2023

## Abstract

We use text-mining techniques to measure the accessibility and quality of information within the texts of interest rate announcements published by the Bank of Israel over the past decade. We find that comprehension of interest rate announcements published by the Bank of Israel requires fewer years of education than interest rate announcements published by the Federal Reserve and the European Central Bank. In addition, we show that the sentiment within these announcements is aligned with economic fluctuations. We also find that textual uncertainty is correlated with the volatility of the domestic financial market.

*Keywords:* Text mining, Central bank communication, Monetary policy, Financial stability.
*JEL Classification:* A12, E44, E52, E58.

---

[*] This paper does not necessarily reflect the views of the Bank of Israel, the Federal Reserve Bank of Richmond or the Federal Reserve System. We thank participants from the Bank of Israel, Harvard University, and the University of Houston research seminars for their useful comments. The replication files of most of the results presented in this paper can be found at doi.org/10.24433/CO.1132001.v1 and central bank text data are available at github.com/JBenchimol/central-bank-texts

[1] Bank of Israel, Research Department, Jerusalem, Israel.
   Corresponding author. Email: jonathan.benchimol@boi.org.il
[2] Federal Reserve Bank of Richmond, Quantitative Supervision and Research, Charlotte, NC, United States.

# 1 Introduction

Being able to properly account for the relationship between central bank communication and financial stability is crucial for policymakers. The Federal Reserve (Fed) aims to build resilience in the financial system and communicate its "policy strategy as clearly and transparently as possible to help align expectations and avoid market disruptions" (Powell, 2018) since information conveyed through central bank communication influences volatility in the financial markets (Benchimol et al., 2020).

In this paper, we explore developments in the readability of announcements published by the Bank of Israel (BoI) throughout the past decades and whether the clarity of written interest rate announcements is related to financial volatility.

Previous studies that analyze central bank policy announcements involve the construction of communication quality indices against which the effect of communication on financial variables (Brand et al., 2010), financial stability (Born et al., 2011), and the future path of interest rates (Bennani et al., 2020) is measured. In this paper, we use similar indices to measure the comprehension and information quality of interest rate announcements published by the Bank of Israel (BoI). We evaluate the extent to which announcement content is (1) accessible to the public; and (2) reflects the economic uncertainty situation and events.[3] We then relate these indices with volatility changes to understand the association between text characteristics and market dynamics.

# 2 Institutional Background

This section describes the main changes that took place in the structure of the BoI's interest rate announcements since understanding them is key to interpreting our quantitative results correctly.

Throughout our sample period, four governors served at the BoI: Jacob Frenkel (1993-2000), David Klein (2000-2005), Stanley Fischer (2005-2013), and Karnit Flug (2013-2018). Nadine Baudot-Trajtenberg served from November 14 to December 23. The structure of the interest rate announcements changed during this period (Fig. 1) and can be divided into three main periods: (1) up to 2006, (2) 2006-2016, and (3) 2016-2019.

In the first period, the length of the interest rate announcement ranged from 200 words to 1000 words and, starting in 2000, included tables with data in addition to the text. The structure and topics discussed were not fixed: the main topics were inflation and the need to bring and maintain it in the target area, the exchange rate and financial stability, and in particular, the consequences of fiscal policy on economic stability. A short paragraph with a central policy message usually appears at the end of the announcement. The texts during this period (2000-2006), being highly redundant and not containing enough words, cannot be used for analyzing communication quality in further sections.

---

[3] Benchimol et al. (2022) review methods for analyzing the text in central bank announcements.

**Figure 1.** Number of Words of the BoI's Interest Rate Announcements

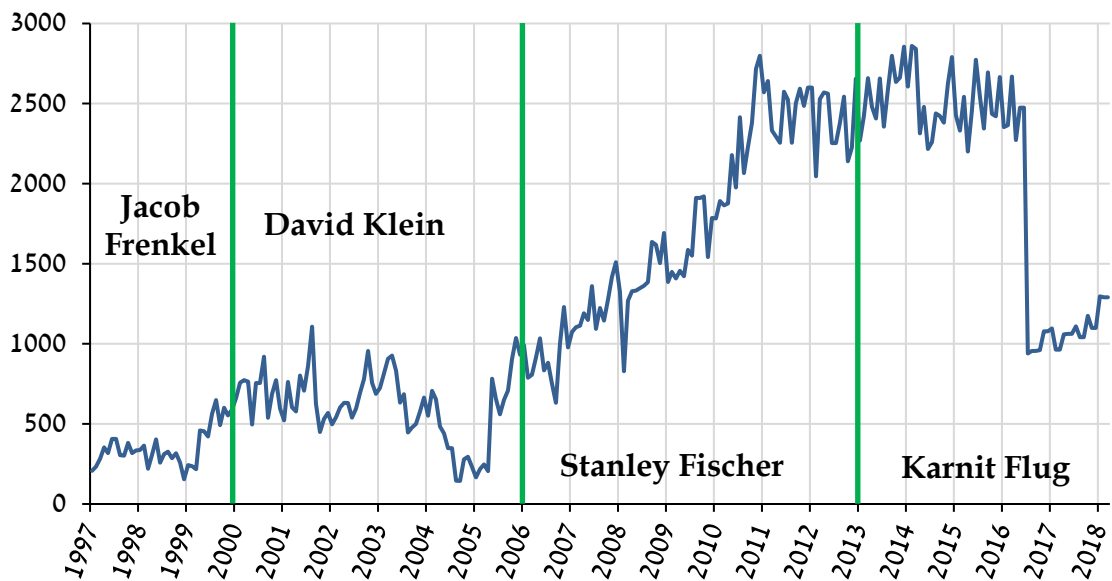

*Source:* Benchimol et al. (2020).

In the second period, the structure of the announcement was divided into two parts: the background conditions included a reference that grew longer over the years to a wide range of economic data: the inflation environment, financial activity, money aggregates, the foreign exchange market, real activity, the global economy, and more. The number of topics discussed expanded according to economic developments. For example, monetary aggregates (M1, M2) gained importance with the global financial crisis (GFC), which led to a paragraph describing their monthly evolution. Against the background of the GFC and the later increases in housing prices, paragraphs describing the developments in the global economy and the housing market were incorporated. The number of words (see Fig. 1) increased from a thousand words to approximately 2500 between 2011 and 2016. The main monetary message was reflected in the main factors for the decision that appeared at the bottom of the announcement, which also included a paragraph expressing a policy statement.

In the third period, against the background of the transition to making interest rate decisions eight times a year (instead of monthly), another change was made to the structure of the interest rate announcements. The main messages appear in summaries on the first page, and the paragraph expressing the policy statement was used to present the forward guidance policy of the monetary policy committee (MPC). The economic developments were mainly presented in the text, as an economic story, with minimal mention of numerical data: these were transferred to a separate file of data that includes a wide variety of charts about the monetary and financial environment, real activity, and the global economy. The change shortened the message significantly, and its length averaged around 1200 words.

Until 2011, the monetary policy was determined by the governor. The governors indeed decided within a forum of managers that resembled the MPC structure, but legally the decision right was reserved for the governor alone. Accordingly, the interest rate announcement reflects the governor's decision (the members of the forum would "vote" in favor of their preferred decision; however, this vote was not binding, and, in some cases, the governor made a different decision than the majority vote). In 2010, the new Bank of Israel law was enacted, and following it, in 2011, the Monetary Committee was established for the first time, consisting of three members from the BoI (the governor, the vice-president, and another employee who is not the governor) and three public representatives. As of October 2011, the interest rate announcement reflects the MPC's decision, which is validated by a majority of votes. The topics highlighted in the announcement, the way they were presented, including the order of appearance and the words were carefully chosen by the monetary decision-makers - the monetary forum followed by the monetary committee. Although complete sentences were often repeated from message to message, there was variation in the degree of detail devoted to the various topics, sub-topics, and wording.

## 3 Communication Quality

The BoI announces interest rate decisions at predetermined dates alongside explanations of the reasoning behind the decisions.[4] For our analysis, we review the content of the English-language interest rate announcements published on the BoI's official website between 2007 and 2018.[5]

### 3.1 Text Accessibility

We measure the accessibility of text within the interest rate announcements using the following two indices. The first one is the type-token ratio (TTR), which captures how the vocabulary used is varied and is calculated as the number of different words (e.g., types) in the text compared to the number of times these words appear (e.g., the number of tokens per type):

$$TTR_t = 100 \left(\frac{Total\ Types_t}{Total\ Tokens_t}\right) \tag{1}$$

---

[4] Several special situations, such as during the GFC, involved interest rate decisions outside the preset dates.
[5] We reasonably assume the nature of the messages the announcements contain is maintained during translation.

A high TTR value indicates the use of a broad vocabulary within the text that requires more significant effort and knowledge to understand.

The second index, text complexity, estimates the number of years of U.S. grade-school-level education required to understand a given text (Kincaid et al., 1975; Flesch, 1979). Since long words and sentences increase communication complexity, the index compares the number of words and sentences with the number of syllables and words,[6] as follows:

$$Complexity_t = 0.39 \left(\frac{Total\ Words_t}{Total\ Sentences_t}\right) + 11.8 \left(\frac{Total\ Syllables_t}{Total\ Words_t}\right) - 15.59 \qquad (2)$$

The initial formula at the origins of Eq. 2, upon which Kincaid et al. (1975) based their results, was developed by Flesch (1948) in an experiment conducted on school-aged children. Using McCall-Crabbs' standard test lessons in reading, Flesch (1948) derived from these experiments correlations between word length (*wl*), sentence length (*sl*), and the average grade of children who could answer at least 50 percent of the test questions correctly.[7] These regressions led to the proposed Reading Ease (RE) formula: RE = 206.835 - .846*wl* - 1.015*sl*. This formula was tested by Flesch (1948) to match various texts and RE results, showing that the correlation between RE and effective RE, measured as the ability of children to understand the suggested texts according to their grade, was significant.

When revising Flesch's RE formula, Kincaid et al. (1975) kept the same predictors but estimated new parameters using a criterion variable that relied on both supply words and multiple-choice test performance. For his experiment, Kincaid et al. (1975) had 531 Navy enlisted personnel take reading comprehension tests called Gates-MacGinitie Reading Test (GMRT) and passages from Rate Training Manuals. Then, multiple regression analysis was conducted using these grades as the criterion variable to validate the revised grade-level version of the Flesch (1948) formula. The Army General Classification Test scores, known for each subject, with the results of multiple regressions, constitute the conversion table from RE (Flesch, 1948) to grade level (Kincaid et al., 1975), represented by the coefficients of Eq. 2.

The outcomes of our analysis reveal that the vocabulary variability (Fig. 2, left panel) rose sharply in 2017 after years of relative stability and, at times, declined. This increase coincided with the change of the announcement format; i.e., shorter texts were used in combination with the provision of more detailed figures about the state of the economy.

Overall, we have 143 documents in our sample, published monthly until January 2017 and eight times a year after, and containing about 2000 words on average.

---

[6] The European Central Bank (ECB) analyzed the readability of their introductory statements with this index (Praet, 2017).

[7] See Table 1 in Flesch (1948).

**Figure 2.** Indices of Variety in Vocabulary in BoI Interest Rate Announcements

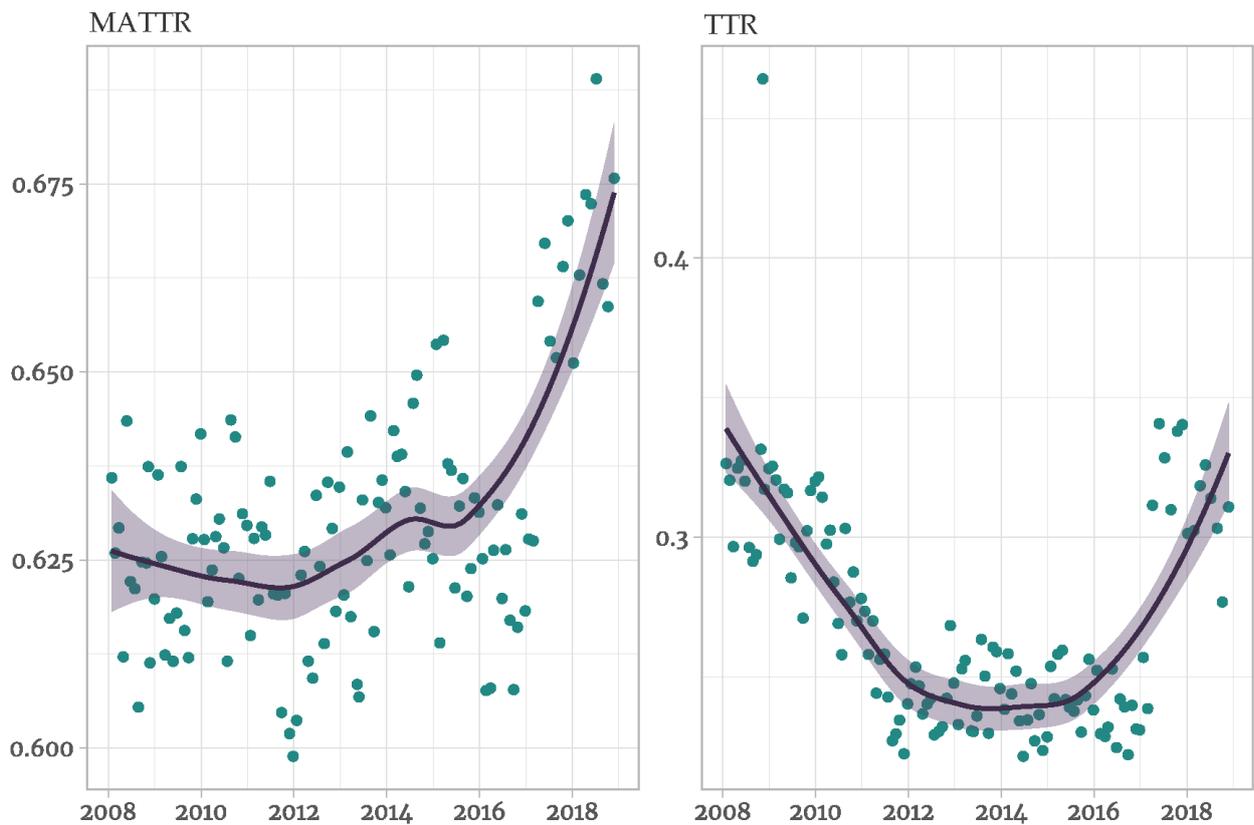

*Notes*: The lines and green dots represent the trend lines and index values, respectively. The TTR is calculated over each announcement. The MATTR is determined over a moving average of 100 words in each announcement.

The TTR index (Fig. 2, right panel) can be sensitive to text length and thus may decline with longer texts.[8] To examine the robustness of our results, we devised a moving average of TTR values in a window that included 100 words (MATTR). Using this approach, a clear increase in text complexity since 2017 can be observed (Fig. 2, left panel).

The Flesch-Kincaid text complexity index (Fig. 3, left panel) exhibits relative stability with slight variance over time. Overall, a U.S. undergraduate-level education (14 years) was required to understand the BoI announcements between 2007 and 2018.

A more precise examination indicated that the average sentence length declined after 2017 (Fig. 3, middle panel) from about 27 to about 22 words per sentence (although levels increased at the end of the sample), making the texts less complex. In contrast, the average number of syllables per word rose (Fig. 3, right panel), increasing text complexity. In sum, the two factors offset each other such that the complexity index remained stable.

---

[8] In general, the more tokens a text contains, the higher the repetition of existing types, particularly punctuation words such as "the" and "and." This leads to an artificial decline in the TTR-defined complexity of the text as sentence length increases (See Eq. 1).

**Figure 3.** Indices of Text Complexity in BoI Interest Rate Announcements

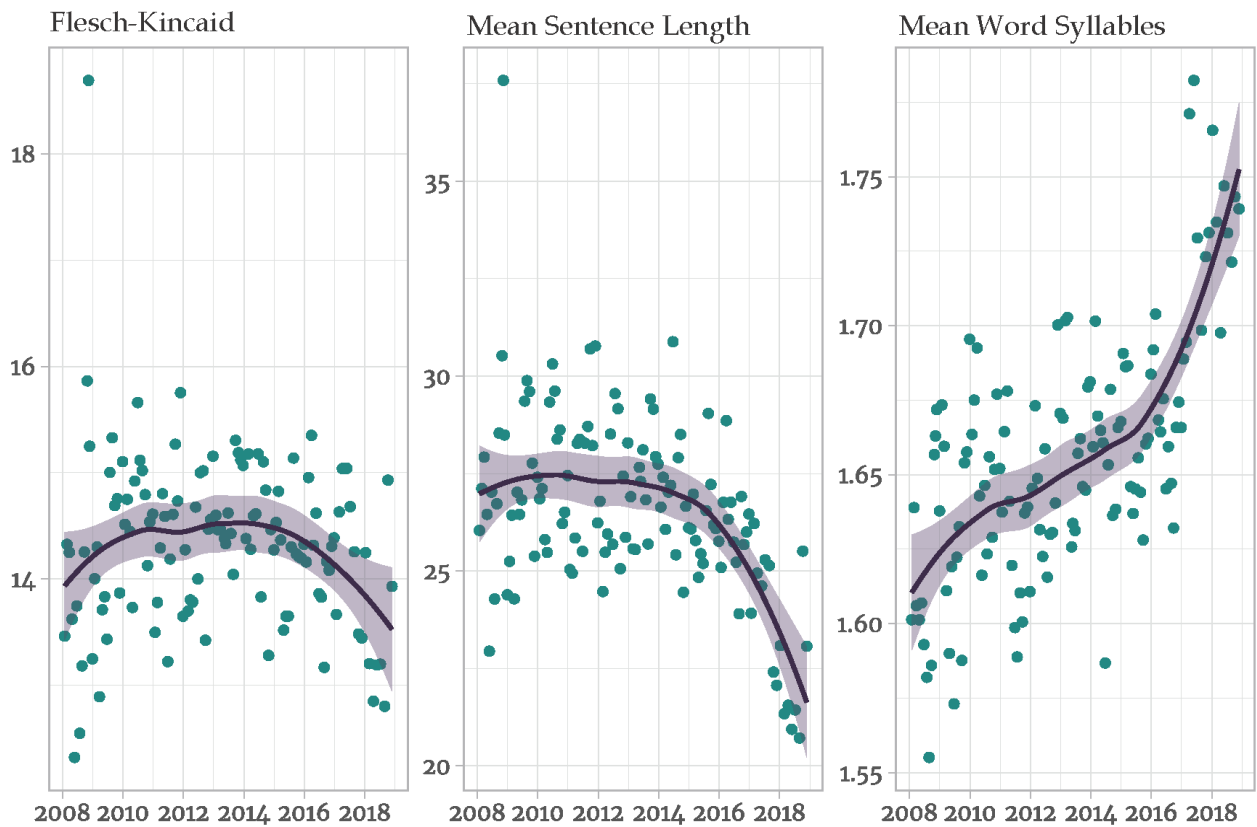

*Note*: The lines and green dots represent trend lines and index values, respectively.

This complexity index does not consider the content of the text and any field-specific knowledge or education that may be required to understand it, regardless of education level. However, it represents an average, and aggregated effects may compensate for heterogeneities. In addition, delivering technical and professional information can also be hampered by excessively low complexity levels.

Governance changes affect interest rate communications differently over time. Under Jacob Frenkel, the length of these communications remained low (Fig. 1). These communications almost doubled in length because of David Klein and the dot-com crisis. Due to the limited number of words and the high amount of redundant text, communication quality is difficult to fully assess before 2008.

With the establishment of the Monetary Policy Committee and support for the new Bank of Israel Law, the Stanley Fischer period marked the most significant changes in these communications. Over the course of his mandate, the length of texts changed from 500 to 2500 words on average (Fig. 1). This increasing dynamic was also influenced by the critical role played by the GFC and financial stability in monetary policy decisions, and did not significantly change lexical diversity (Fig. 2) or readability (Fig. 3). However, this dynamic increased the number of syllables per word (Fig. 3), indicating a slight increase in text complexity.

During the mandate of Karnit Flug, the BoI reduced the frequency of interest rate decisions from monthly to eight times a year. This change involved a drastic decrease in the interest rate communication length, switching from an average of 2500 words to an average of 1200 words per interest rate announcement (Fig. 1), which also contributed to the increase in lexical diversity (Fig. 2) and complexity (Fig. 3). Still, overall communication quality was hardly affected as these texts were more accessible (Flesc-Kincaid declined in Fig. 3) and sentence length was reduced (see the mean sentence length, Fig. 3).

The changes in governors with the governance of historical (dot-com and GFC) and legal contexts (Bank of Israel Law and the establishment of the Monetary Committee) influenced the BoI's communications.

**Figure 4.** International Comparison of Text Complexity Indices and Linguistic Variety

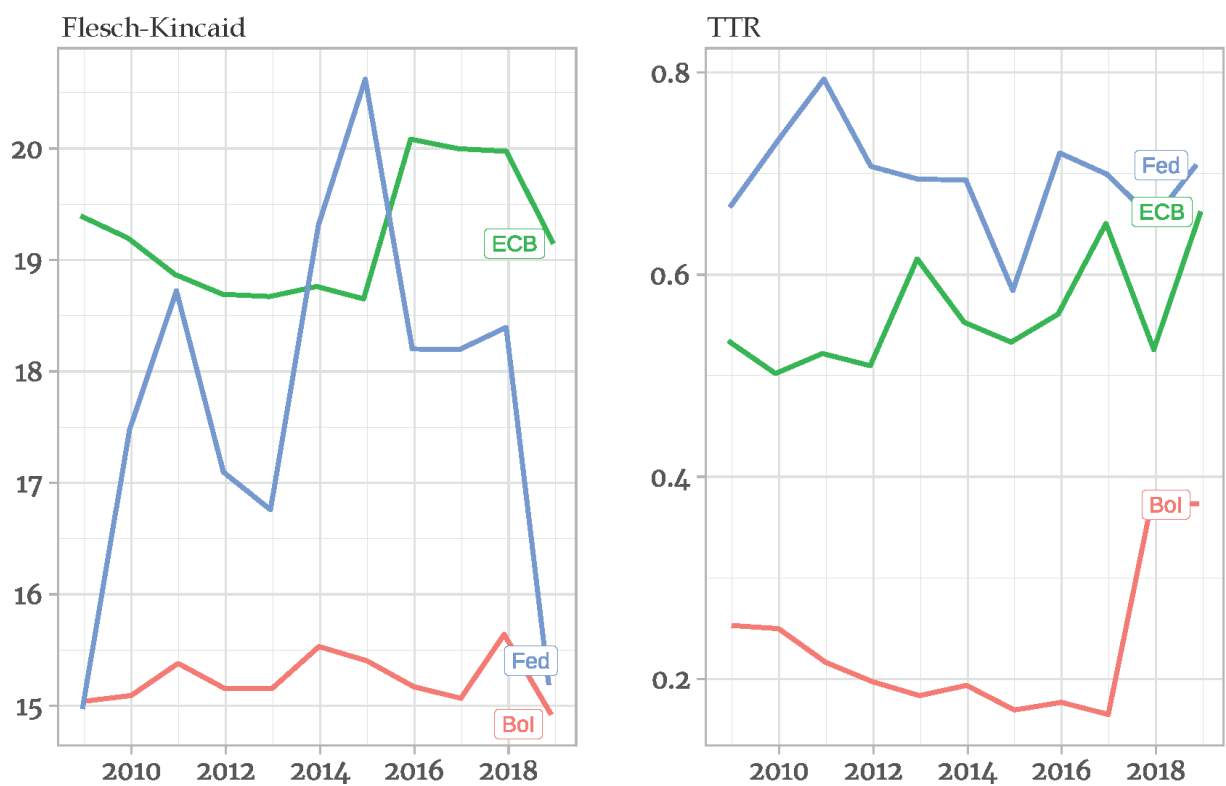

*Note:* Yearly averages based on interest rate announcements.
*Source:* Benchimol et al. (2020).

To interpret these indices objectively, we look at the interest rate announcements made by other central banks. Fig. 4 compares the complexity and linguistic variety indices of the BoI's announcements with the indices of the communications published by the Fed and the ECB. The Fed and ECB's announcements require an average of about 17 and 19 years of education, respectively, to be understood. The BoI announcements are relatively more comprehensible, requiring an average of about 14 years of education to be understood. Furthermore, the Fed and the ECB's linguistic variety indices are almost three times as high as that of the BoI on average.

## 3.2 Contextual Uncertainty

In this section, we examine the quality of the information in the interest rate announcements based on the financial context of the words they contain.[9] Instead of analyzing divergence in tone (Sadique et al., 2013; Tilleman and Walter, 2019), we build a textual uncertainty index. In the first stage, using standard dictionary-based methods that are designed to analyze financial texts, we identified words in the text that conveyed uncertainty;[10] for example, "risk," "uncertainty," "volatility," probability," and "variable."

**Figure 5.** Index of Uncertainty in the BoI Interest Rate Announcements and TA-35 VIX

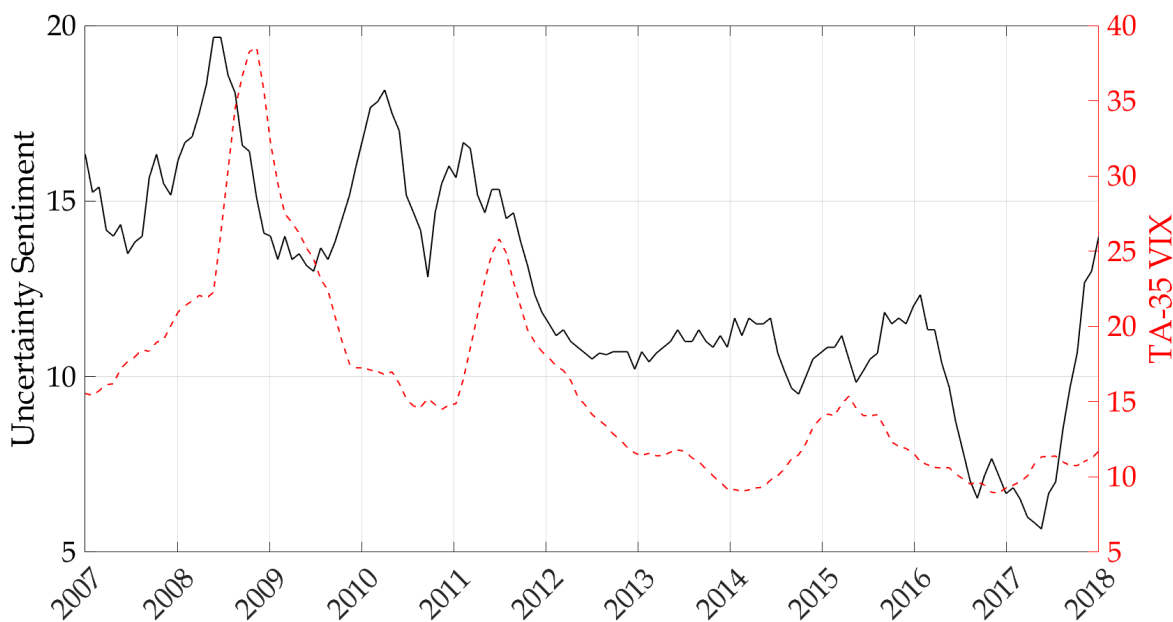

*Source:* Bank of Israel and Benchimol et al. (2020).

In the second stage, we calculate an index of textual uncertainty, which increases as the rate of words classified in the category of uncertainty increases. The uncertainty index (Fig. 5) shows that the announcements contained relatively high contextual uncertainty around the time of the GFC. The contextual uncertainty significantly declined after that, particularly after 2016, before rising in response to the global economic uncertainties that transpired toward the end of the sample period.[11]

---

[9] Information quality and uncertainty conveyed are different concepts and should be distinguished. The BoI could express a high degree of uncertainty when actual uncertainty is indeed high, which is high-quality information. Whether the text is of high or low quality depends on its signal-to-noise ratio.
[10] The classification adopted is based on the categories proposed by Loughran and McDonald (2011).

The literature on dictionary-based classifications typically employs a unigram approach, which examines individual words independently. Our methodology aligns with this approach, acknowledging that it may lead to inaccuracies in instances where a two-term phrase, composed of independently negative words, is incorrectly classified as negative when it conveys a positive meaning (e.g., "uncertainty declined"). Despite this potential limitation, the unigram approach is well-suited for our specific corpus of relatively short texts, the structural characteristics of these texts (redundancies), and the dictionaries utilized in previous studies. Additionally, central bank announcements tend to avoid using double-negative structures and compensations between false positives and false negatives may be at play. Also, using a bigram or phrase-level classification would significantly decrease the information present in short texts, such as interest rate announcements.

We note several instances of correlation between our uncertainty index and real-world economic events. We also examine the correlation between uncertainty indices and risk levels in domestic markets, as reflected in the TA-25 VIX index.[12] A positive Pearson correlation of 0.47 was determined across the sample, while a distance correlation of 0.54 confirmed both linear and nonlinear linkages—causality direction or the assertion that uncertainty leads to fluctuations in VIX, or vice versa, could not be inferred from the data available. The announcements may contain analyses the public is unaware of, causing market volatility. In contrast, the text may describe VIX fluctuations that would have occurred in any case, even without the announcement. Further in-depth research is required to identify causality direction.[13]

## 4 Conclusion

Our findings, namely that the BoI's announcements on interest rates are easier to read than those issued by the ECB or Fed and convey meaningful signals about what is happening in the economy, are encouraging. In addition, the changes made in the interest rate announcement structure and voting procedure over the past decade did not negatively affect the readability of the text and its signal.

---

[11] See the Global Economic Policy Uncertainty Index (Baker et al., 2016).
[12] This index is calculated using the implied volatility of options on the index. The TA-25 index was changed to the TA-35 index in February 2017.
[13] Our uncertainty index positively correlates with the U.S. VIX index, although it is lower (Pearson: 0.44; Distance: 0.50). This finding, together with the fact the BoI's interest rate announcements do not influence the U.S. VIX, supports the hypothesis that volatility in the markets causes an increase in the level of uncertainty in the BoI interest rate announcements, not the other way around.

We interpret these results as evidence that complexity and signal do not trade-off. Making the interest rate announcement more readable or less complex does not result in a decrease in the quality of economic information.

Sharing transparent communications with the public has become an essential central bank tool that significantly contributes to financial stability. The findings of this study indicate that understanding the BoI's interest rate announcements requires a college-level education, albeit a considerably lower level of education than that needed to understand the Fed and the ECB announcements.

Furthermore, we show that introducing a new announcement format at the beginning of 2017 has made the texts less comprehensible. Finally, by deriving a text uncertainty index from the BoI's interest rate announcements, we find a correlation between these announcements and economic events and preliminary evidence that contextual uncertainty correlates with changes in the TA-25 VIX index.